\documentclass[12pt]{article}

\pdfoutput=1
\usepackage[margin=0.75in]{geometry}
\usepackage{textcomp}
\usepackage{listings}
\usepackage{xcolor}
\usepackage[hyphens]{url}
\usepackage{hyperref}
\hypersetup{
    colorlinks,
    linkcolor={red!50!black},
    citecolor={blue!50!black},
    urlcolor={blue!80!black}
}

\title{SimpleSBML: A Python package for creating, editing, and interrogating SBML models: Version 2.0}

\author{ Herbert M. Sauro\footnote{to whom correspondence should be addressed}\\ Department of Bioengineering,\\University of Washington, Seattle, WA 98195, USA}

\date{September 2, 2020 -- Updated with corrections August, 2021}

\definecolor{keywords}{RGB}{255,0,90}
\definecolor{comments}{RGB}{0,0,113}
\definecolor{red}{RGB}{160,0,0}
\definecolor{green}{RGB}{0,150,0}
\begin{document}
\maketitle
\noindent\Large\textbf{Abstract}

\noindent\normalsize
In this technical report, we describe a new version of SimpleSBML which provides an easier to use interface to python-libSBML allowing users of Python to more easily construct and inspect SBML based models. The most commonly used package for constructing SBML models in Python is python-libSBML based on the C/C++ library libSBML. python-libSBML is a comprehensive library with a large range of options but can be difficult for new users to learn and requires long scripts to create even the simplest models. Inspecting existing SBML models can also be difficult due to the complexity of the underlying object model. Instead, we present SimpleSBML, a package that allows users to add and inspect species, parameters, reactions, events, and rules to a libSBML model with only one command for each.  Models can be exported to SBML format, and SBML files can be imported and converted to SimpleSBML commands that create each element in a new model.  This allows users to create new models and edit existing models for use with other software. In the new version, a range of `get' methods is provided that allows users to inspect existing SBML models without having to understand the underlying object model used by libSBML. 

\noindent\large\textbf{Accessibility and Implementation:}

\noindent\normalsize
SimpleSBML is publicly available and licensed under the liberal MIT open source license.  It supports SBML levels 2 and 3.  Its only dependency is libSBML.  It is supported on Windows, Mac OS, and Linux. All code has been deposited at the GitHub site \url{https://github.com/sys-bio/simplesbml} and is available for user installation via a standard command: {\tt pip install simplesbml}. From environments such as Spyder (\url{https://www.spyder-ide.org/}), one an install packages by simply typing {\tt pip install package-name} at the python console. In a Jupyter notebook one needs to escape the pip command by using: {\tt !pip install package-name}, note the exclamation point. 

\noindent\large\textbf{Contact:}

\noindent\normalsize
\noindent \url{hsauro@uw.edu}

\noindent\large\textbf{Supplementary information:}

\noindent\normalsize
User documentation is available at \url{sys-bio.github.io/simplesbml} as well as a pip install on pypi. 

\section{Introduction}

In an earlier report~\cite{Cannistra030312} we described SimpleSBML, a python library that makes it much easier to create SBML models. In this document, we describe a new version of simpleSBML that extends the functionality of simpleSBML to now include a series of {\tt get} methods to easily extract information from an existing SBML model.

Biological modeling is a key component of systems biology, and the development of Systems Biology Markup Language (SBML)~\cite{sbml}, a markup language designed to describe models of biological systems, has allowed systems and synthetic biologists to develop a plethora of useful software tools that are automatically compatible with each other through SBML document import and export.  Creating SBML models is possible with a variety of software packages, the most well-known being libSBML~\cite{libsbml}. This package allows users to generate SBML documents by writing scripts in Python, C or C++.  While libSBML is very useful, it can be difficult to learn for novices and even the simplest of models using libSBML tend to be very long.  SimpleSBML is a package that allows users to create SBML models with Python scripting, but requires far fewer commands and is more accessible for beginners. One other difference between version 1 and 2 of SimpleSBML is that the class name has been capitalized to match common practice used in Python code, hence we will use {\tt SbmlModel} instead of {\tt sbmlModel}.

SimpleSBML makes use of libSBML methods to add and inspect elements such as species, parameters, reaction, events and rules to a model, and generate an SBML-formatted document from the resulting model object.  SimpleSBML also includes two new methods that allow users to load existing SBML models either from a string or file. SimpleSBML can therefore be used to edit SBML-formatted models as well as create new ones.

\section{Methods}

Verison 1.0 of simplesbml is given in the previous publication~\cite{Cannistra030312} and the examples for creating a new SBML model will not be repeated other than giving one example for reference. The SbmlModel class, holds a SBMLDocument object and contains methods that can be used to add different elements, such as species, parameters, reactions and events.  Here is an example of a simple reaction-based model built with SbmlModel.

\begin{lstlisting}[language=Python]
import simplesbml
model = simplesbml.sbmlModel();
model.addCompartment(1e-14, comp_id='comp');
model.addSpecies('E', 5e-21, comp='comp');
model.addSpecies('S', 1e-20, comp='comp');
model.addSpecies('P', 0.0, comp='comp');
model.addSpecies('ES', 0.0, comp='comp');
model.addReaction(['E', 'S'], ['ES'], 'comp*(kon*E*S-koff*ES)', \
        local_params={'koff': 0.2, 'kon': 1000000.0}, rxn_id='veq');
model.addReaction(['ES'], ['E', 'P'], 'comp*kcat*ES', \
		local_params={'kcat': 0.1}, rxn_id='vcat');
\end{lstlisting}

\subsection{New {\tt get} API}

Of more interest is the new {\tt get} API that allows an existing SBML model to be easily inspected. This section will discuss the new {\tt get} API that verison 2.0 has. The first change is to allow existing SBML models to be loaded into simpleSBML. The existing constructor {\tt SBMLModel} was modified to accept SBML strings and file names containing SBML. To make the interface simpler to use, two additional methods outside the class, {\tt loadSBMLStr} and {\tt loadSBMLFile} are provided. For example, to load a file one can use the following code:

{\tt model = simplesbml.loadSBMLFile ('mymodel.xml')}

This returns an instance of {\tt SbmlModel}. The online documentation gives a full list of methods that are available but a comment on the naming and class structure philosophy is in order. To begin with, the entire API is flat, that is there are no subclasses which one finds, for example, in libsbml. This means a user does not need to know the underlying object model in order to effectively use the API. However to make this work, the names for the API methods must be clear in what they do and can be identified using intellisense and code completion in IDEs (Integrated Development Environment) such as spyder (\url{https://www.spyder-ide.org/}). For example, to obtain the list of reaction Ids in a model requires the following code when using libsbml:

\begin{lstlisting}[language=Python]
nReactions = document.model.getNumReactions() 
for i in range (nReactions):
    p = document.model.getReaction(i)
    print (p.getId())             
\end{lstlisting}  

where {\tt document} was obtained by calling the libsbml method: 

{\tt document = reader.readSBMLFromString(sbmlStr)}. 

This requires three levels of indirection as well as knowledge of the various methods to call at each level. Instead, simplesbml allows a user to call a single method:

\begin{lstlisting}[language=Python]
  print (model.getListOfReactionIds())
\end{lstlisting}  

The method name is qualified with the term {\tt Ids} because one could also retrieve the SBML names rather than the Ids. The entire API is built along these lines. As an illustration, the code below will use simplesbml to construct the stoichiometric matrix for a model. Tellurium~\cite{telluriumChoi} is used to convert the reaction scheme in antimony~\cite{antimony} format into SBML which is then loaded into simplesbml.

\begin{lstlisting}[language=Python]
import tellurium as te, simplesbml,  numpy as np

r = te.loada("""
ext S4  # S4 is a boundary species

S0 + S3 -> S2; k0*S0*S3;
S3 + S2 -> S0; k1*S3*S2;
S5 -> S2 + S4; k2*S5;
S0 + S1 -> S3; k3*S0*S1;
S5 -> S0 + S4; k4*S5;
S0 -> S5; k5*S0;
S1 + S1 -> S5; k6*S1*S1;
S3 + S5 -> S1; k7*S3*S5;
S1 -> S4 + S4; k8*S1;

S0 = 0; S1 = 0; S2 = 0; S3 = 0; S4 = 0; S5 = 0;
k0 = 0; k1 = 0; k2 = 0; k3 = 0; k4 = 0
k5 = 0; k6 = 0; k7 = 0; k8 = 0
""")

model = simplesbml.loadSBMLStr(r.getSBML())

stoich = np.zeros((model.getNumFloatingSpecies(), model.getNumReactions()))
for i in range (model.getNumFloatingSpecies()):
    floatingSpeciesId = model.getNthFloatingSpeciesId (i)
    
    for j in range (model.getNumReactions()):
        productStoichiometry = 0; reactantStoichiometry = 0

        numProducts = model.getNumProducts (j)
        for k1 in range (numProducts):
            productId = model.getProduct (j, k1)

            if (floatingSpeciesId == productId):
               productStoichiometry += model.getProductStoichiometry (j,k1)

        numReactants = model.getNumReactants(j)
        for k1 in range (numReactants):
            reactantId = model.getReactant (j, k1)
            if (floatingSpeciesId == reactantId):
               reactantStoichiometry += model.getReactantStoichiometry (j,k1)

         stoich[i,j] = int(productStoichiometry - reactantStoichiometry)
print (stoich)
\end{lstlisting}  

The resulting stoichiometry matrix is given by:

\begin{lstlisting}[language=Python]
[[-1.  1.  0. -1.  1. -1.  0.  0.  0.]
 [-1. -1.  0.  1.  0.  0.  0. -1.  0.]
 [ 1. -1.  1.  0.  0.  0.  0.  0.  0.]
 [ 0.  0. -1.  0. -1.  1.  1. -1.  0.]
 [ 0.  0.  0. -1.  0.  0. -2.  1. -1.]]
\end{lstlisting}  

\section{Installation}

SimpleSBML depends on python-libSBML so that libSBML must be installed. Refer to the~\url{https://pypi.python.org/pypi/python-libsbml} for details. SimpleSBML can installed using the standard pip mechanism for Python packages by using the command:

\medskip
{\tt pip install simplesbml}

\medskip
When using more sophisticated Python environments such as Jupyter (\url{https://jupyter.org/}), Spyder or PyCharm (\url{https://www.jetbrains.com/pycharm/}), it is possible to type {\tt pip install simplesbml} at the Python console itself and the package will be installed automatically. If you're using a more basic Python console such as Idle, you will need to use the pip command from the operating system terminal. Make sure that Python and pip are on the search path if you take this approach. On Windows the easiest way to get simplesbml is to install the Tellurium modeling package found at \url{http://tellurium.analogmachine.org/}. This comes ready bundled with simplesbml. 

\section{Testing and Documentation}

Documentation is generated from comments in the source code using sphinx and uploaded to readthedocs. Additional documentation is provided by including text in the {\tt index.rst} file. 

Testing was originally done using the Python unittest package (\url{https://docs.python.org/3/library/unittest.html}) but was found to have limited capabilities. The missing functionality included an easy way to test the API on a variety of loaded SBML models and to exercise the API more thoroughly. Such a workflow is not supported well by unittest. Instead pyTest (\url{https://docs.pytest.org/en/stable/}) was tried which does support parameterized tests but the code became too verbose and unwieldy. Instead a very simple testing framework was written which was ultimately easier to add new tests to. The test file is called {\tt runTests.py} and is located in the {\tt tests} folder. To run the tests, simple run this file from Python.

\section{Discussion}

SimpleSBML is intended for systems biology researchers who have limited experience with programming, or are working on simple models and prefer to use a simpler set of commands compared to libSBML. Future additions to the software may include additional methods to deal with model annotations and the SBML layout and render standard.

\section*{Acknowledgments}

\paragraph{Funding:} The authors are most grateful to generous funding from the National Institute of General Medical Sciences of the National Institutes of Health under award R01GM-123032 as well as the NSF award CCF-0432190. The content is solely the responsibility of the authors and does not necessarily represent the official views of the National Institutes of Health, the National Science Foundation, or the University of Washington. I thank Joe Hellerstein, Lucian Smith and Ciaran Welsh for useful discussions.


\begin{thebibliography}{9}

\bibitem{sbml}
Hucka, M., A. Finney, H. M. Sauro, et al. (2003) The Systems Biology Markup Language (SBML): A Medium for Representation and Exchange of Biochemical Network Models. \emph{Bioinformatics} 19(4): 524?531, doi:10.1093/bioinformatics/btg015.


\bibitem{libsbml}
    Bornstein, B.J., 
    Keating, S. M.,
    Jouraku, A.,
    and Hucka M. (2008)
    LibSBML: An API library for SBML.
    \emph{Bioinformatics}, 24(6):880-881,
    doi:10.1093/bioinformatics/btn051.
  
\bibitem{libroadrunner}
    Somogyi, E.T,
    Bouteiller, J-M,
    Glazier, J.A.,
    K{\"o}nig, M,
    Medley, K,
    Swat, M.H,
    Sauro, H.M. (2015) libRoadRunner: A High Performance SBML Simulation and Analysis Library.
    Bioinformatics. 2015 Oct 15;31(20):3315-21. https://doi.org/10.1093/bioinformatics/btv363

\bibitem{Cannistra030312}
	Cannistra, Caroline,
	Medley, Kyle,
	Sauro, Herbert M., (2015),
	SimpleSBML: A Python package for creating and editing SBML models, doi = {10.1101/030312}, bioRxiv

\bibitem{telluriumChoi}
   Choi, K., Medley, J. K., König, M., Stocking, K., Smith, L., Gu, S., and Sauro, H. M. (2018). Tellurium: An extensible python-based modeling environment for systems and synthetic biology. Bio Systems, 171, 74–79. https://doi.org/10.1016/j.biosystems.2018.07.006

\bibitem{antimony}
Smith, L. P., Bergmann, F. T., Chandran, D., and Sauro, H. M. (2009). Antimony: a modular model definition language. Bioinformatics (Oxford, England), 25(18), 2452–2454. https://doi.org/10.1093/bioinformatics/btp401

 \end{thebibliography}
\end{document}